\documentclass[a4paper]{article}

\usepackage{INTERSPEECH2022}
\usepackage{stfloats}

\title{SASV Based on Pre-trained ASV System and Integrated Scoring Module}
\name{Yuxiang Zhang$^{12}$, Zhuo Li$^{12}$, Wenchao Wang$^1$, Pengyuan Zhang$^{12}$}
\address{
  $^1$Key Laboratory of Speech Acoustics and Content Understanding, Institute of Acoustics, China\\
  $^2$University of Chinese Academy of Sciences, Beijing, China}
\email{\{zhangyuxiang, lizhuo, wangwenchao, zhangpengyuan\}@hccl.ioa.ac.cn}

\begin{document}

\maketitle
\begin{abstract}
	Based on the assumption that there is a correlation between anti-spoofing and speaker verification, a Total-Divide-Total integrated Spoofing-Aware Speaker Verification (SASV) system based on pre-trained automatic speaker verification (ASV) system and integrated scoring module is proposed and submitted to the SASV 2022 Challenge. The training and scoring of ASV and anti-spoofing countermeasure (CM) in current SASV systems are relatively independent, ignoring the correlation. In this paper, by leveraging the correlation between the two tasks, an integrated SASV system can be obtained by simply training a few more layers on the basis of the baseline pre-trained ASV subsystem. The features in pre-trained ASV system are utilized for logical access spoofing speech detection. Further, speaker embeddings extracted by the pre-trained ASV system are used to improve the performance of the CM. The integrated scoring module takes the embeddings of the ASV and anti-spoofing branches as input and preserves the correlation between the two tasks through matrix operations to produce integrated SASV scores. Submitted primary system achieved equal error rate (EER) of 3.07\% on the development dataset of the SASV 2022 Challenge and 4.30\% on the evaluation part, which is a 25\% improvement over the baseline systems. 
\end{abstract}

\section{Introduction}
As a type of biometric technology, automatic speaker verification (ASV) \cite{reynolds1995speaker} has made significant progress in recent years. From early algorithms based on statistical machine learning \cite{reynolds2000speaker, kenny2005joint, dehak2010front} to current deep learning based models \cite{variani2014deep, snyder2018x, desplanques20_interspeech}, ASV systems with increasingly lower equal error rates (EERs) are gradually being used in practice. However, there are various spoofing attacks against ASV systems, including impersonation, replay, text-to-speech (TTS) and voice conversion (VC). These spoofing attacks can cause serious performance degradation of ASV systems \cite{wu2015spoofing}.

To protect ASV systems from spoofing attacks, the biennial ASVspoof challenges were successfully held from 2015 to 2021 \cite{wu2015asvspoof, Kinnunen2017, todisco2019asvspoof, yamagishi21_asvspoof}. Based on the datasets provided by the challenges, a large number of countermeasures (CM) with good performance against TTS and VC have emerged \cite{lavrentyeva2019stc, wang21fa_interspeech, tak21_asvspoof, ge21_asvspoof}. Even in complex scenarios such as codecs and channel mismatch, the CMs can still achieve good results \cite{tomilov21_asvspoof, chen21b_asvspoof}. However, these CMs are built independently of ASV systems, and they are only aimed at anti-spoofing, not in conjunction with ASV systems. The ASVspoof challenges provide the minimum normalized tandem detection cost function (min t-DCF) \cite{kinnunen2020tandem} metric to simulate and evaluate the performance of anti-spoofing systems in tandem with ASV systems. But the metric still has significant limitation, as the ASV system is fixed.

Few attempts have been made to jointly optimize ASV systems with CM systems. Spoofing-Aware Speaker Verification (SASV) systems can be divided into joint-decision based methods and multi-task based methods depending on whether ASV and CM systems are trained independently. Joint-decisions can be made from the feature level such as i-vectors and embeddings obtained from ASV system and CM respectively \cite{sizov2015joint, gomez2020joint}. Another approach of joint-decision making is fusion of scores through different back-end processing methods \cite{sahidullah2016integrated, todisco2018integrated, shim2020integrated, kanervisto2021optimizing}. The joint-decision based methods need to train the ASV and CM systems separately, with less scope for joint optimization. The multi-task based methods make ASV system and CM share some feature extraction layers and optimize the two tasks simultaneously \cite{li2019multi, li2020joint, zhao2022multi}. However, multi-task based methods consider less about the correlation between two tasks when outputting fusion results. Therefore, it is of great significance to study the correlation between ASV and anti-spoofing tasks in terms of features and scoring to obtain an integrated ASV system with anti-spoofing capability.

In order to provide a unified framework and evaluation metrics to support the development of single integrated SASV systems, as well as to improve robustness, the SASV Challenge 2022 \cite{jung2022sasv} has been held as a special session of INTERSPEECH 2022. In the SASV Challenge 2022, only logical access (LA), i.e., TTS and VC spoofing attacks, are considered. 

In this paper, based on the assumption of the correlation between the ASV task and the anti-spoofing task, a Total-Divide-Total (TDT) integrated SASV system based on pre-trained ASV subsystem and integrated scoring module is proposed. The features in the time delay neural network (TDNN) based pre-trained ASV system \cite{desplanques20_interspeech} are demonstrated can be used for LA spoofing speech detection. Especially the features after multi-layer feature aggregation (MFA) can get the optimal results. These features are fed into the subsequent dual-branch network for ASV and spoofing speech detection, respectively. In addition, the performance of CM is demonstrated can be further improved by concatenating the embeddings extracted from two branches. The two embeddings are then fed into an integrated scoring module to obtain the SASV scores. The integrated scoring module is also based on the assumption of correlation and is implemented through the operations of the score matrix. The proposed integrated SASV system outperformed baseline systems on the SASV-EER metric by adding several layers to the pre-trained ASV subsystem based on the correlation.

\section{Total-Divide-Total SASV architecture}
The submitted systems are all based on the ECAPA-TDNN pre-trained baseline ASV subsystem provided by challenge organizers. Firstly, the method of anti-spoofing based on the pre-trained ASV system is introduced. The integrated TDT SASV structure and the integrated scoring module are described bellow.

\subsection{Anti-spoofing with pre-trained ASV system}\label{CM}
The SASV challenge provided two baseline systems for the SASV 2022 development and evaluation partitions. Based on the ECAPA-TDNN pre-trained baseline ASV subsystem, an anti-spoofing approach is proposed.

There are two reasons motivating us to try to obtain an anti-spoofing CM through features extracted from the given pre-trained ASV subsystem. On the one hand, the state-of-the-art (SOTA) ASV subsystem performed well on the speaker verification (SV) task, achieving an EER of 0.83\% on the evaluation set. Despite the poor performance of 29.32\% EER on the anti-spoofing task, this indicated that the embeddings of ECAPA-TDNN pre-trained ASV system still have some ability to discriminate spoofing speech. In other words, this suggested a correlation between the SV task and the anti-spoofing task. On the other hand, through sharing feature extraction layers from the pre-trained ASV model, the same features can be used as the SV task. This approach is consistent with the concept of an integrated system, and it allows the correlation between the two tasks to be maintained and fully exploited. 

The concatenated features obtained from the MFA of the ECAPA-TDNN pre-trained ASV system are fed into one of the subsequent branches of the dual-branch network for spoofing speech detection. In addition, the speaker embeddings extracted by the fixed SV branch are used to assist in the optimization of the CM. A linear layer is added after the CM embedding concatenated with the speaker embedding in an attempt to improve the performance of the CM branch by utilizing the speaker information, as well as to enhance the correlation.

\subsection{Total-Divide-Total structure}
\begin{figure}[t]
	\centering
	\includegraphics[width=\columnwidth]{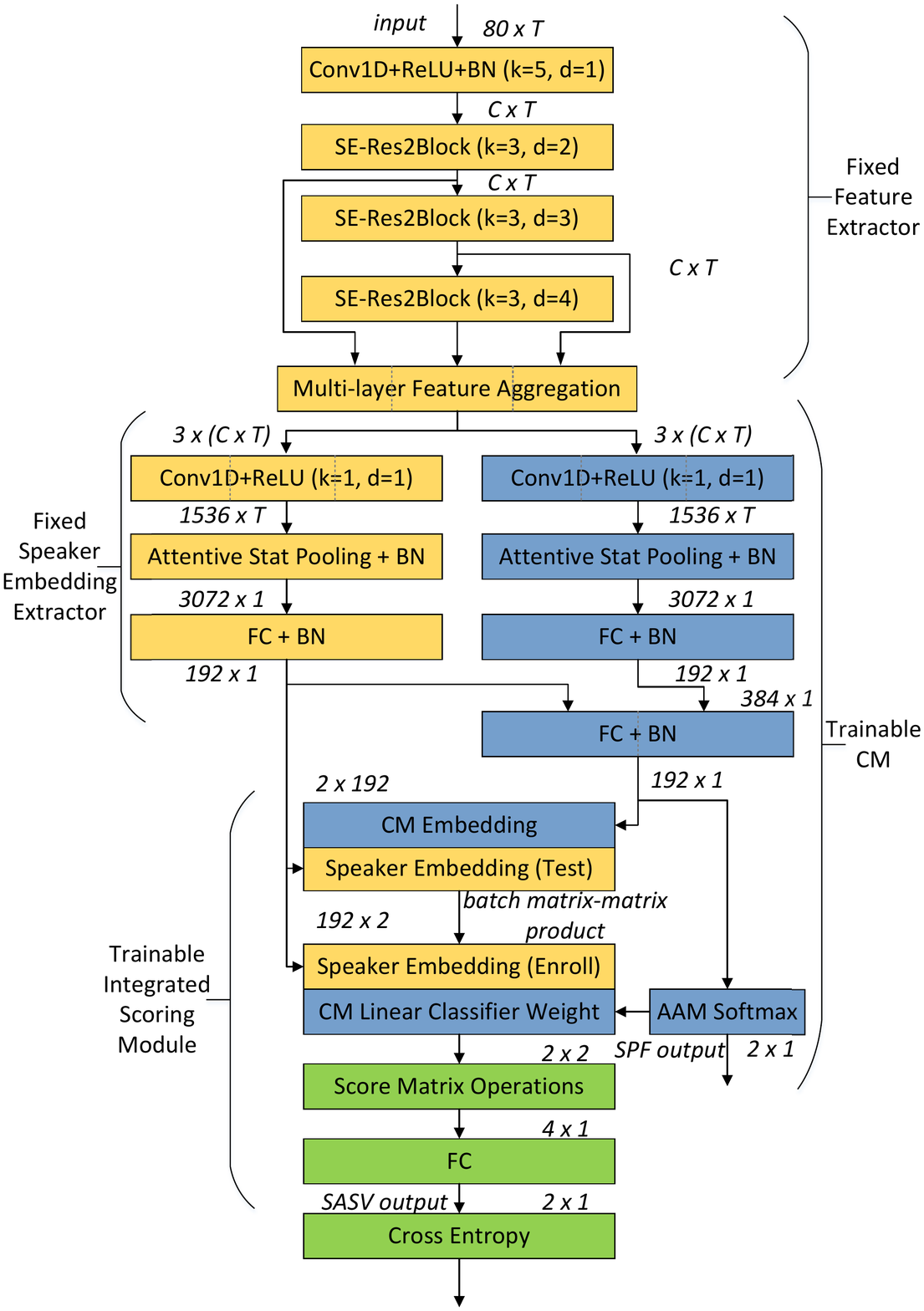}
	\caption{Network topology of the Total-Divide-Total ECAPA-TDNN pre-trained ASV system based SASV structure .}
	\label{fig:tdt}
\end{figure}

As shown in Figure \ref{fig:tdt}, the ``Total-Divide-Total'' SASV structure consists of a total feature extraction network, a divided dual-branch network for ASV and anti-spoofing embedding extraction, as well as a total integrated scoring module.

The ``Total-Divide'' part is similar to the multi-task based SASV system. However, in the dual-branch structure, the SV branch is fixed and only the anti-spoofing branch is optimized. The reasons are the excellent performance on the SV task of the pre-trained ASV system and the large time consumption required to train the ASV model.

After yielding the final ASV embeddings and CM embeddings separately, they are not used for scoring individually, but are fed into the integrated scoring module described below.

\subsection{Integrated scoring module}
Two integrated scoring strategies were used in our submitted systems. The first strategy is score integration through score matrix operations. Suppose that the test speaker embedding and the CM embedding are $\bm{E_{test}}$ and $\bm{E_{CM}}$ respectively, while the enroll speaker embedding is $\bm{E_{en}}$ and the weight of the last classifier in CM is $\bm{W}$. Note that the final linear classification layer of CM is a $2\times192$-dimensional matrix, and the final score can be obtained by the difference between the two scores: $S_1-S_0$. So the difference of the vectors obtained by $\bm{W}=\bm{W}_1-\bm{W}_0$ is used as the scoring vector. Then the score matrix $\bm{S}$ can be obtained from the following equation:
$$
\bm{S}=
\begin{bmatrix}
	\eta_1 & S_{SV} \\
	S_{CM} & \eta_2
\end{bmatrix}_{2\times2} = 
\left[
\begin{array}{c}
	\bm{E_{test}} \\
	\bm{E_{CM}} \\
\end{array}
\right]_{2\times 192}
\left[
\begin{array}{c}
	\bm{W}\\ \bm{E_{en}}
\end{array}
\right]_{2\times 192}^T
$$

Since we argue that there is a correlation between embeddings from ASV branch and embeddings from spoofing speech detection branch, the elements on the diagonal of the score matrix $\bm{S}$ are maintained. Similar to \cite{zhang2022new}, the score matrix $\bm{S}$  is transformed into a probability matrix $\bm{P}$ by the Sigmoid function:
\begin{eqnarray*}
	\bm{P}=\sigma(\bm{S})=
	\begin{bmatrix}
		\theta_1 & P_{SV}\\
		P_{CM} & \theta_2
	\end{bmatrix}
\end{eqnarray*}
where $\sigma$ denotes the sigmoid function.

In order to derive the probability of the bona-fide speech belonging to the target speaker, inspired by \cite{zhang2022new} and the Baseline1, the probability matrix is then simply manipulated so as to obtain both the sum and the product of the probabilities:

\begin{equation*}
	\begin{split}
		\bm{J}&=\bm{P}+\bm{P}^T+\bm{P}\cdot\bm{P}\\
		&=
		\begin{bmatrix}
			\delta_{1}+P_{SV}P_{CM} & \epsilon_1 P_{SV}+\epsilon_2 P_{CM} \\
			\epsilon_1 P_{CM}+\epsilon_2 P_{SV} & \delta_{2}+P_{SV}P_{CM}
		\end{bmatrix}
	\end{split}
\end{equation*}
where $\delta_{1}=\theta_1^2+2\theta_1$,  $\delta_{2}=\theta_2^2+2\theta_2$, $\epsilon_1=1+\theta_1$, $\epsilon_2=1+\theta_2$, contain the information of correlation.

After obtaining the probability matrix $\bm{J}$, it is flattened into a vector and fed into a linear layer, thus completing the final binary classification.

Another integrated scoring strategy is simpler. A matrix with shape of $5\times192$ is obtained by concatenating the five vectors mentioned above: $\bm{E_{test}}, \bm{E_{en}}, \bm{E_{CM}}, \bm{W}_1, \bm{W}_2$. Then, the final $192$-dimentional embedding is obtained by a two-dimensional convolutional layer with a kernel of $5\times1$ and a convolutional layer with $1\times1$ kernel.

\section{Experimental setup}

\subsection{Dataset and metrics}
All experiments presented further were performed on ASVspoof 2019 LA dataset \cite{wang2020asvspoof}. Only the training set was used to train all systems. The development part was used for performance validation.

The metrics of the challenge are classical EERs, including SASV-EER as the primary metric, SV-EER for SV task, and EER estimates spoofing attacks (SPF-EER). The SASV-EER does not distinguish between different speaker (zero-effort, non-target, or impostor) access attempts and spoofed access attempts. The SV-EER is traditional estimates of speaker verification performance evaluated from a set of target and non-target trials. While the SPF-EER is similar to SV-EER except that non-target trials are replaced with spoofing trials.

\subsection{Details of systems implementation}\label{sys}
Except for the TDT architechture and the intrgrated scoring module, the parameters of the convolutional layers of the proposed model is identical to the ECAPA-TDNN pre-trained system. The network leverages some of the latest advances in deep learning to achieve SOTA performance on the VoxCeleb1-O \cite{nagrani2017voxceleb} dataset. The input features are the 80-dimensional FBank. Since SpecAug has no obvious improvement on ASVspoof 2019 LA dataset in our past experiments, the same data augmentation method except SpecAug is used here. 

During the training process, the same Additive Angular Margin Softmax \cite{deng2019arcface} is used to optimize the CM. The SASV task is optimized with binary cross entropy loss with weights of $(0.1, 0.9)$ which is the same as in the Baseline2.

The details of five single systems are as follows:
\begin{itemize}
	\item TDT-1 and TDT-2: Two implementations of the proposed TDT system described above. The integrated scoring strategy is based on the score matrix operations.
	\item TDT-A: Similar to the TDT-1, but with learnable weights obtained by simple attention (A) layer multiplied to the two embeddings before calculating the score matrix.
	\item TDT-O: Similar to the TDT-1, but with the addition of a loss function to make the two weight vectors of the last classification layer of CM as orthogonal (O) as possible, thus enhancing the robustness of the CM branch.
	\item TDT-C: Similar to the TDT-1, but the integrated scoring strategy is based on convolution (C).
\end{itemize}

The primary system submitted to the challenge is a fusion of five single systems at the score level. The fusion is done with the same weights of $0.2$.

\section{Results and Discussion}
\subsection{Results for anti-spoofing with pre-trained ASV system}

\begin{table}[th]
	\caption{Results of the countermeasures based on pre-trained ASV system on the ASVspoof 2019 LA development and evaluation subset in terms of EER and min t-DCF.}
	\label{cm}
	\centering
	\begin{tabular}{ l c c c r}
		\toprule
		& \multicolumn{2}{c}{min t-DCF} & \multicolumn{2}{c}{EER/\%} \\
		\cmidrule{2-5}
		& Dev & Eval & Dev & Eval\\
		\midrule
		CM1 & 0.0218 & 0.1047 & 0.75 & 3.64 \\
		CM2 & 0.0094 & 0.0775 & \textbf{0.31} & 2.79 \\
		CM3 & \textbf{0.0085} & \textbf{0.0680} & 0.39 & \textbf{2.09} \\
		CM4 & 0.1378 & 0.1545 & 4.59 & 5.97 \\
		\bottomrule
	\end{tabular}
\end{table}

Results of pre-trained ASV system based CM on ASVspoof 2019 LA development and evaluation dataset are presented as EER and min t-DCF. We tried training from different layers, and the results are shown in the Table \ref{cm}. The systems from the CM1 to the CM4 stand for: all parameters are fully trainable; trainable from the layer before the MFA; trainable after the MFA; only the last linear layer can be trained.

The results from the Table \ref{cm} show that the pre-trained ASV system can be used for the anti-spoofing task. On this basis, spoofing speech detection and SV task can be integrated more adequately. The EER of the best CM is close to 2\%. Different performance can be obtained by freezing different layers, so it is important to select the right starting layer. All parameters in CM1 can be trained, but the results are only better than that of CM4, which training only the last linear layer. Making the layers below the MFA trainable and freezing the layers above, CM3 gets the best results. CM2 has a lower EER on the Dev set compared to CM3, but a higher one on the Eval set, thus the CM2 with one more block trained is slightly overfitted.


\begin{table}[th]
	\caption{Results in terms of three EERs of speaker embedding assisted countermeasure on the SASV 2022 development and evaluation subset. CM3-A stands for CM3 trained with the assistance (A) of speaker embedding features.}
	\label{assist}
	\centering
	\begin{tabular}{ l c c c c c r}
		\toprule
		& \multicolumn{2}{c}{SV-EER/\%} & \multicolumn{2}{c}{SPF-EER/\%} & \multicolumn{2}{c}{SASV-EER/\%} \\
		\cmidrule{2-7}
		& Dev & Eval & Dev & Eval & Dev & Eval \\
		\midrule
		CM3 & 34.23 & 46.16 & 0.34 & 1.92 & 13.27 & 24.36 \\
		CM3-A & 32.43 & 45.06 & 0.41 & \textbf{1.78} & 12.96 & 23.58 \\
		\bottomrule
	\end{tabular}
\end{table}

The experimental results in the Table \ref{assist} demonstrate the effect of the speaker embeddings extracted by the pre-trained ASV system to assist in the optimization of the CM branch. Through concatenating the embeddings of the CM with the speaker embeddings extracted by the pre-trained ASV system and adding a linear layer, the performance of CM branch on SASV metrics is improved. All three EERs are reduced slightly. In particular, the SPF-EER decreases from 1.92\% to 1.78\%, a relative decrease of 7.3\%. This suggests that, due to the correlation between the anti-spoofing task and the SV task, the speaker information contained in the embeddings extracted by the pretrained ASV system is helpful for the anti-spoofing task, especially in the SASV system.

\subsection{Ablation study of the integrated scoring module}\label{ism}

\begin{table}[th]
	\caption{Results in terms of three EERs of different integrated scoring strategies on the SASV 2022 datasets.}
	\label{score}
	\centering
	\begin{tabular}{ l c c c c c r}
		\toprule
		& \multicolumn{2}{c}{SV-EER/\%} & \multicolumn{2}{c}{SPF-EER/\%} & \multicolumn{2}{c}{SASV-EER/\%} \\
		\cmidrule{2-7}
		& Dev & Eval & Dev & Eval & Dev & Eval \\
		\midrule
		TDT-B & 10.99 & 27.08 & 1.13 & 5.20 & 4.21 & 14.34 \\
		TDT-C & 13.60 & 6.28 & 1.35 & 7.50 & 6.80 & 7.08 \\
		TDT-M & 4.58 & 9.21 & 0.47 & 2.20 & 3.57  & 7.03 \\
		TDT-D & 4.18 & 6.95 & 1.31 & 2.19 & 2.83 & 5.08 \\
		TDT-1 & 3.44 & \textbf{6.24} & 0.37 & \textbf{1.73} & 2.76 & \textbf{4.78} \\
		\bottomrule
	\end{tabular}
\end{table}

To study the effectiveness of different scoring methods in the integrated scoring module, five different integrated scoring strategies are explored. In addition to TDT-1 and TDT-C introduced in Section \ref{sys}, TDT-B in this section represents the method of concatenating the three embeddings similar to the Baseline2 (B). TDT-D represents the method of setting the diagonal (D) elements of the score matrix to zero, based on the assumption that the CM and ASV are independent of each other. TDT-M represents multiplying (M) the probability of CM by the probability of ASV based on the independence assumption.

As the results in the Table \ref{score} show, the performance of the systems optimized by the score obtained from dot product of embeddings is better than that of the systems optimized directly by embedding concatenation or convolution. The worst results are obtained by concatenating the embeddings, losing the well performance of the pre-trained ASV model on the SV task. The convolution-based scoring strategy is better than embedding concatenation, but still worse than the three strategies based on scores. This is because the convolution operation overemphasizes the correlation of different embeddings. However, the difference between the three EERs is relatively small compared to TDT-B on the evaluation set. Among systems based on the dot product and independence assumption, TDT-D with integrated scoring by score matrix operation works better than TDT-M using probability multiplication directly. Although the SV-EER of TDT-M on the development set is almost the same as that of TDT-D, and the SPF-EER was 0.84\% lower, TDT-D has better generalization. All three EERs of TDT-D outperforms TDT-M on the evaluation set. This indicates that the integrated scoring strategy based on the score matrix operations is more suitable for SASV through obtaining both sum and product of probabilities at the same time. The optimal results are obtained by the proposed approach of preserving the correlation between CM and SV branches through the score matrix operations in TDT-1, with metrics outperforming other scoring strategies on both the development and evaluation sets.

\subsection{Results for Total-Divide-Total SASV systems}

\begin{table*}[th]
\caption{Results in terms of three different EERs on the SASV 2022 development and evaluation subsets.}
\label{results}
\centering
\begin{tabular}{ l c c c c c r}
	\toprule
	& \multicolumn{2}{c}{SV-EER/\%} & \multicolumn{2}{c}{SPF-EER/\%} & \multicolumn{2}{c}{SASV-EER/\%} \\
	\cmidrule{2-7}
	& Dev & Eval & Dev & Eval & Dev & Eval \\
	\midrule
	Baseline1 (score-sum) \cite{jung2022sasv} & 32.88 & 35.32 & 0.06 & 0.67 & 13.07 & 19.31 \\
	Baseline2 (back-end ensemble model) \cite{jung2022sasv} & 12.87 & 11.48 & 0.13 & 0.78 & 4.85 & 6.37\\
	\midrule
	TDT-1 & 3.44 & \textbf{6.24} & 0.37 & 1.73 & 2.76 & \textbf{4.78} \\
	TDT-2 & 3.64 & 6.81 & 0.39 & 2.35 & 2.76 & 5.14\\
	TDT-A (weighted embeddings) & 10.65 & 8.44 & 0.73 & 2.59 & 5.05 & 6.17 \\
	TDT-O (add constraints) & 6.20 & 8.85 & 0.43 & \textbf{1.49} & 4.31 & 7.02\\
	TDT-C (convolution based scoring) & 13.60 & 6.28 & 1.35 & 7.50 & 6.80 & 7.08 \\
	\midrule
	Fusion with equal weights & 4.25 & \textbf{5.62} & 0.20 & \textbf{1.21} & 3.07 &  \textbf{4.30}\\
	\bottomrule
\end{tabular}
\end{table*}

The results of single and fusion TDT SASV systems based on pre-trained ASV system and integrated scoring module on the SASV 2022 dataset are shown in the Table \ref{results}. Results confirme the efficiency of the proposed TDT SASV system. The SASV-EER obtained by TDT-1 and TDT-2 single systems outperforms the Baseline2 on both the development and the evaluation datasets. The best single system TDT-1 obtaines SASV-EER of 4.78\% on the evaluation set, almost 25\% reduction compared to the Baseline2 system. The EER of the fusion system on the evaluation set is 4.30\%, which was a further 10\% reduction compared to the best single system. 

Comparing the results of different single systems, adding mechanisms such as attention to TDT systems could lead to performance degradation, especially for SV-EER. A higher SV-EER is obtained because the SV branch is fixed. The weights given to the two embeddings by attention further increased this gap. For the TDT-O system, the lowest single system SPF-EER on the evaluation set is obtained by constraining the orthogonality of the weights of the classification layer. Maintaining the orthogonality of the classification layer weights can improve the performance of the anti-spoofing system and enhance the robustness of the embeddings of the CM branch. As discussed in Section \ref{ism}, the convolution-based integrated scoring strategy is inferior in all aspects compared to the score matrix operations based strategies, but more balanced.

It also can be found that the SPF-EER is much lower compared to the SV-EER in all systems, since only the spoofing speech detection branch and the integrated scoring module were optimized. Compared to the large gap between SV-EER and SPF-EER in the two baseline systems, the two metrics of the proposed TDT systems have a smaller gap due to the utilization of the correlation and thus performs better on the SASV-EER metric. However, the proposed result ranks 18th, which is a big gap compared to other SOTA systems \cite{jung2022sasv1}. It remains to be investigated whether fine-tuning the pre-trained ASV system while optimizing the CM and SASV can provide better results.

\section{Conclusion}
This paper describes the Total-Divide-Total integrated SASV system based on pre-trained ASV system and integrated scoring module submitted to the SASV 2022 Challenge. According to the assumption that there is a correlation between the ASV and the anti-spoofing tasks, based on the finding that the features from the pre-trained ASV system can be utilized for spoof speech detection, an integrated SASV system can be obtained by training several more layers based on the pre-trained ASV system. An integrated scoring module based on score matrix operations is also proposed, which also preserves the correlation between the two tasks and naturally yields an integrated SASV score. The proposed SASV system achieved significant improvements compared to the baseline systems.

\section{Acknowledgements}
This work is partially supported by the National Key Research and Development Program of China (No. 2021YFC3320103).

\bibliographystyle{IEEEtran}

\bibliography{mybib}

\end{document}